\documentclass[a4paper,fleqn,usenatbib]{mnras}

\usepackage[T1]{fontenc}
\usepackage{ae,aecompl}

\usepackage{graphicx}	% Including figure files
\usepackage{amsmath}	% Advanced maths commands
\usepackage{amssymb}	% Extra maths symbols

\def\deg{{^\circ}}

\def\nsim{{\sim\!}}
\title[Inner Galactic Stellar Halo]{The Stellar Halo in the Inner Milky Way: Predicted Shape and Kinematics}

\author[P\'erez-Villegas, Portail \& Gerhard] 
{Angeles P\'erez-Villegas$^{1}$ \thanks{E-mail: mperez@mpe.mpg.de}, Matthieu Portail$^{1}$ and Ortwin Gerhard$^{1}$
\\
$^{1}$ Max-Planck-Institut f\"ur Extraterrestrische Physik, Gie\ss enbachstra\ss e, D-85741 Garching, Germany\\    
}

\date{Submitted 4 July 2016}
\pubyear{2016}

\begin{document}
\label{firstpage}
\pagerange{\pageref{firstpage}--\pageref{lastpage}}
\maketitle

\begin{abstract}
  We have used N-body simulations for the Milky Way to investigate the kinematic and structural properties of
  the old metal-poor stellar halo in the barred inner region of the Galaxy. We find that the extrapolation of the
  density distribution for bulge RR Lyrae stars, $\rho\sim r^{-3}$, approximately matches the number density of RR Lyrae
  in the nearby stellar halo. We follow the evolution of such a tracer population through the formation and evolution of
  the bar and box/peanut bulge in the N-body model. We find that its density distribution changes from oblate to
  triaxial, and that it acquires slow rotation in agreement with recent measurements. The maximum radial velocity is
  $\sim15-25$ km/s at $\vert l\vert\!=10^\circ-30^\circ$, and the velocity dispersion is $\sim120$ km/s.  Even
  though the simulated metal-poor halo in the bulge has a barred shape, just $12\%$ of the orbits follow the bar, and it
  does not trace the peanut/X structure. With these properties, the RR Lyrae population in the Galactic bulge is
  consistent with being the inward extension of the Galactic metal-poor stellar halo.
\end{abstract}

\begin{keywords}
Galaxy: bulge -- Galaxy: Kinematics and dynamics -- Galaxy: structure --Galaxy: centre -- methods: numerical
\end{keywords}

%%%%%%%%%%%%%%%%%%%%%%%%%%%%%%%%%%%%%%%%%%%%%%%%%%

%%%%%%%%%%%%%%%%% BODY OF PAPER %%%%%%%%%%%%%%%%%%

\section{Introduction}\label{Intro}

The inner region of the Milky Way (MW) hosts multiple components such as the bar and box/peanut (B/P) bulge, the thin and
thick disks, and the inner stellar halo. Stars in the Galactic bulge thus occupy a wide range in the Metallicity
Distribution Function (MDF), with [$\rm{Fe/H}$] between $-3.0$ and $+1.0$ dex \citep{Rich88, Zoccali2003, Ness2013a,
  Gonzalez2015}.

In the bulge region of the Galaxy a small but not insignificant fraction of metal-poor stars has been detected. About
$5\%$ of the ARGOS sample \citep{Ness2013a} have metallicities $[\rm{Fe/H}]<-1.0$. These metal-poor stars show a slower
rotation and a higher velocity dispersion than the stars with metallicity $[\rm{Fe/H}]> -1.0$ \citep{Ness2013b}. Their
different kinematics is possible evidence that the metal-poor bulge stars are not part of the B/P bulge, but belong to a
distinct component.  Other studies conclude that the MDF in the bulge can be separated into a metal-rich population of
stars that are part of the B/P bulge, and a metal-poor population that traces an old spheroidal component
\citep{Babus2010, Hill2011,Rojas-Are2014}.

RR Lyrae stars (RRLs) are well-known tracers of old, metal-poor populations, and have been used to trace the old
component in the Galactic bulge.  The bulge RR Lyrae population has a MDF centered around $[\rm{Fe/H}]=-1.0$, with a
small spread in metallicity \citep{Walker91,Pietru2012}.  Thousands of RRLs have been discovered in the bulge
\citep{Soszy11, Soszy14}, and these stars are ideal tracers of structures because their distances can be accurately
estimated. However, there is disagreement on the spatial distribution of the bulge RRLs: \citet{Dekany2013} find that
the RRL distribution is spheroidal with a slight elongation inside 1 kpc, whereas, \citet{Pietru2015} find
that the spatial distribution of RRLs is barred, but that the RRLs are not part of the X-shaped structure seen in red
clump giant stars \citep[e.g.][]{Wegg2013}.  The old RRL component in the bulge could represent the inner extension of
the Galactic halo \citep{Alcock98,Minniti99} even though it is barred, due to gravitational effects of the bar that
formed later in the disk \citep{Saha12}. A recent radial velocity (RV) study in the inner bulge ($\vert l\vert<4\deg$)
has shown that the RRL rotate slowly if at all \citep{Kunder2016}.

Combining these new data, detailed studies can now be made of the properties of the stellar halo in the inner region
around the Galactic bar.  The questions that we want to address in this study are: Could the metal-poor stars in the
bulge be part of the inner stellar halo? What is the influence of the bar and B/P bulge on the kinematics and spatial
distribution of these stars? We use N-body simulations to address these questions because they are powerful tools to study bars and bulges in the evolution of
galaxies \citep{Combes90,Athana02,Debattista06}, and similarly the bulge/bar region of the MW \citep{ Fux97, Sevenster99,Shen2010,Martinez-Valpuesta11, Portail15}. In this paper, Section \ref{sec:density} discusses the density profile of MW
RRLs. Section \ref{sec:model} describes the N-body method that we use for modelling the stellar halo in the bar
region. Our predictions for the shape and kinematics of the inner stellar halo are presented in Section
\ref{sec:prediction}. Section \ref{sec:orbits} contains an orbital analysis of the stellar halo particles. Our
conclusions are presented in Section \ref{sec:conclusions}.

% It is important to mention that the peak of the MDF for RRLys far away from the Galactic center is more metal-poor,
% with $[\rm {Fe/H}]\sim -1.5$ (Torrealba et al 2015, Juric et al 1996).

\section{Density Profile of RRLs}\label{sec:density}

\begin{figure}
\vspace{-0.4cm}
\includegraphics[width=\columnwidth]{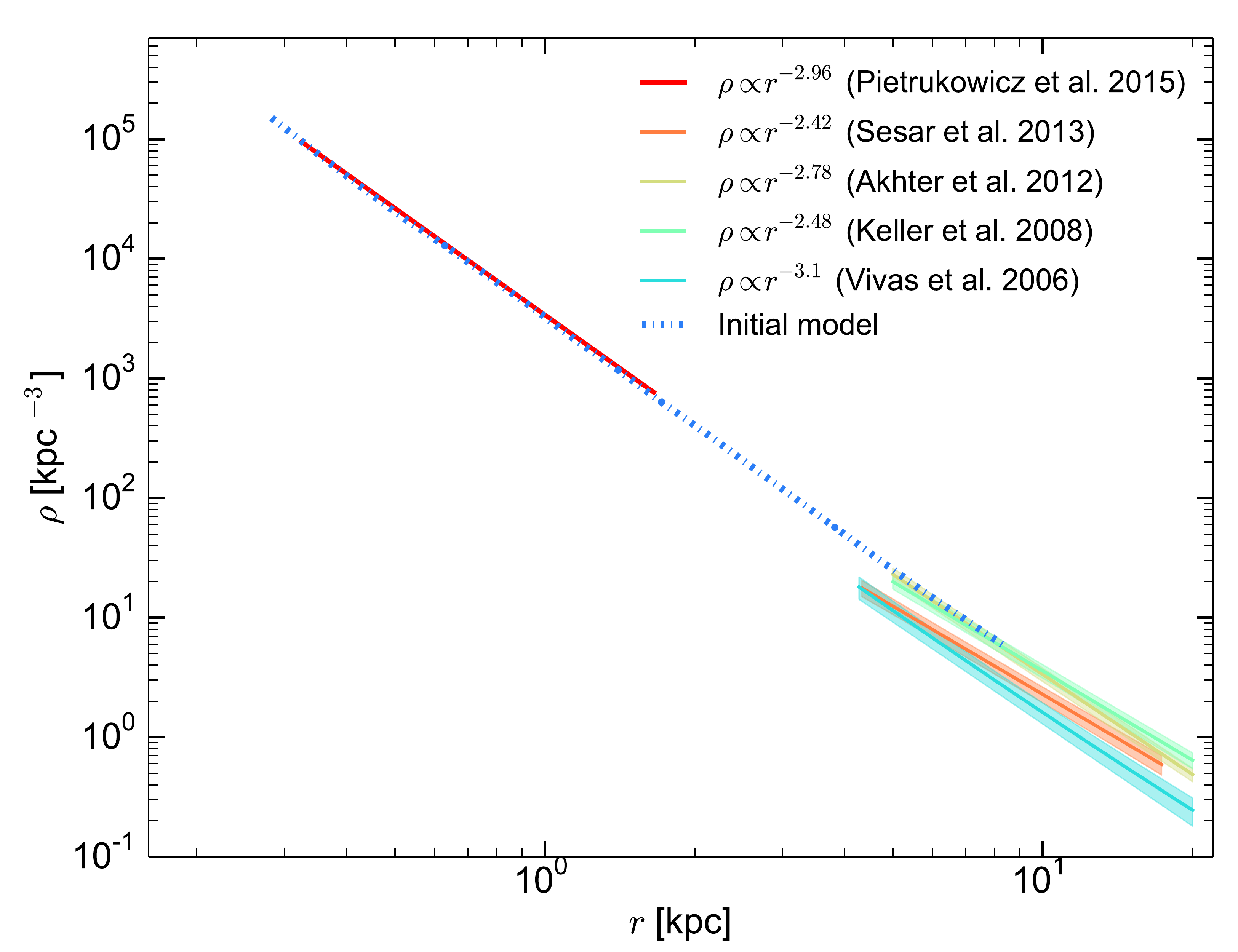}
\vspace{-0.4cm}
\caption{RRL spatial density profiles as function of spherical radius. The red solid line shows the deprojected density profile for bulge RRLs determined from the surface density profile given in \citet{Pietru2015}. The coloured solid lines between $\sim 5-20$
  kpc radius show the density profiles of stellar halo RRLs from the studies indicated, the shaded areas show the $1\sigma$ errors. The dotted line is the stellar halo density profile of the initial model described in Section \ref{sec:model}.}
\label{fig:Den_bul_SH}
\end{figure}

The density profile of the RRLs in the stellar halo between radii of $5-20$ kpc is often described as a single power
law, $\rho \sim r^{\alpha}$, with slope $\alpha$ between $-2.4$ and $-3.1$
\citep{Wett96,Amrose01,Vivas06,Keller08,Juric08,Akhter12,Sesar13}. Could the bulge RRL be the inner extension
of the RRL halo?

Figure \ref{fig:Den_bul_SH} shows the deprojected density profile of bulge RRLs\footnote{ This corrects  the normalization in eq. (17) of \citet{Pietru2015}.} calculated from the surface density profile given by \citet{Pietru2015} as function of spherical radius $r$, defined for a triaxial model as $r=\sqrt[3]{abc}$, if $a,b,c$ are the axis lengths of the given isodensity contour. The density profile of the bulge RRLs is well described by a power law with a $\alpha \nsim -2.96$ in the distance range between $0.2-2.8$ kpc.  Also shown in Fig.~\ref{fig:Den_bul_SH}
are density profiles for stellar halo RRLs from the literature, for radii between $\sim5-20$ kpc. We see that if we extrapolate the power-law density profile for the bulge RRLs to larger radii, it matches the density normalizations of the halo RRLs within a factor $\sim 2$. Based on this argument, bulge and stellar halo RRLs could be part of the same component.

We note that the MDF of the bulge RRLs has a peak at $[{\rm Fe/H}]=-1.0$ \citep{Pietru2015} whereas the halo RRL population
peaks at $[{\rm Fe/H}]=-1.4$ \citep{Jurcsik96, Nemec2013, Torrealba2015}. In our interpretation, this implies a metallicity
gradient in the bulge-halo RRL population. This could arise, for example in an accretion model for the halo, because the fragments that reached the centre of the MW were chemically somewhat more evolved than those that contributed most of the stars near the Sun.  
Further information about the history of the halo could be provided by the $\alpha$-enhancement . For RRLs in the halo $[\alpha/ {\rm Fe}]\sim 0.3$ dex \citep{Nissen2010,For2011}, however for RRLs in the bulge there are no measurements of $[\alpha/ {\rm Fe}]$ available in the literature yet. In future, it will be interesting to compare such measurements with the halo RRLs.

Therefore, henceforth we will consider the hypothesis that the RRLs in the bulge are the inner
extension of the stellar halo.

%\begin{equation}
% \label{eq:density}
%\rho(m)= \rho_0 \left( \frac{m}{R_0} \right)^{-3},
%\end{equation}
%$\rho_0$ is the density of the RRLs at the Sun position, and $R_0$ is the position of the Sun.

\section{Stellar halo model: N-body simulation} \label{sec:model}

We assume that the metal-poor halo traced by RRLs formed early in the history of the MW before the Galactic bar was
present.  At these early times, we model the stellar halo as an oblate component with axial ratios $b/a=1.0$, $c/a =0.6$ and a
single power-law density profile $\rho\sim m^{-3}$, where
$m=\sqrt{x^2+\left(\frac{y}{b/a}\right)^2+\left(\frac{z}{c/a}\right)^2}$. The assumed flattening is consistent with
values observed near the Sun \citep{Morrison_etal00, Siegel_etal02, Chen_etal01, Phleps_etal05, Sesar13} and the normalization is chosen to agree with the bulge RR Lyare profile in Fig.~\ref{fig:Den_bul_SH}. This model is shown by the dotted line in Fig. \ref{fig:Den_bul_SH}.

To follow the stellar halo through the bar and buckling evolution of the MW, we use the N-body simulation M85 from
\citet{Portail15} as it evolves in its self-consistent potential $\Phi(\mathbf{x},t)$, from the initial time
($t\!=0$) through bar formation ($1.6$ Gyr) and buckling ($2.8$ Gyr) to its final time ($5.2$ Gyr).  We use all the dark
matter halo particles from model M85 as test particles which now will be representing the stellar halo. The dark matter particles now have two weights, the first is to calculate the dark matter potential, and the second is the weight that orbits have in the stellar halo. We determine the stellar halo
weights using the Made-to-Measure method \citep{SyT96, deLorenzi07}, such that at $t\!=0$ they represent
our oblate power-law density model for the stellar halo in $\Phi(\mathbf{x},t\!=0)$. Then at any later time $t$ we
can use these same particle weights to determine the properties of the stellar halo in the evolved
$\Phi(\mathbf{x},t)$. Note that our initial stellar halo does not rotate, and has a mild radial anisotropy.

\section{Predicted Shape and Kinematics of the Inner Stellar Halo} \label{sec:prediction}

Now we use the halo particle weights to reconstruct the stellar halo at the final time of the simulation,
thereby predicting the shape of its density distribution and the kinematics of the halo stars after several Gyr of
gravitational interaction with the bar and B/P bulge.

\subsection{Shape of the stellar halo}\label{sec_shape}

\begin{figure}
\includegraphics[width=\columnwidth]{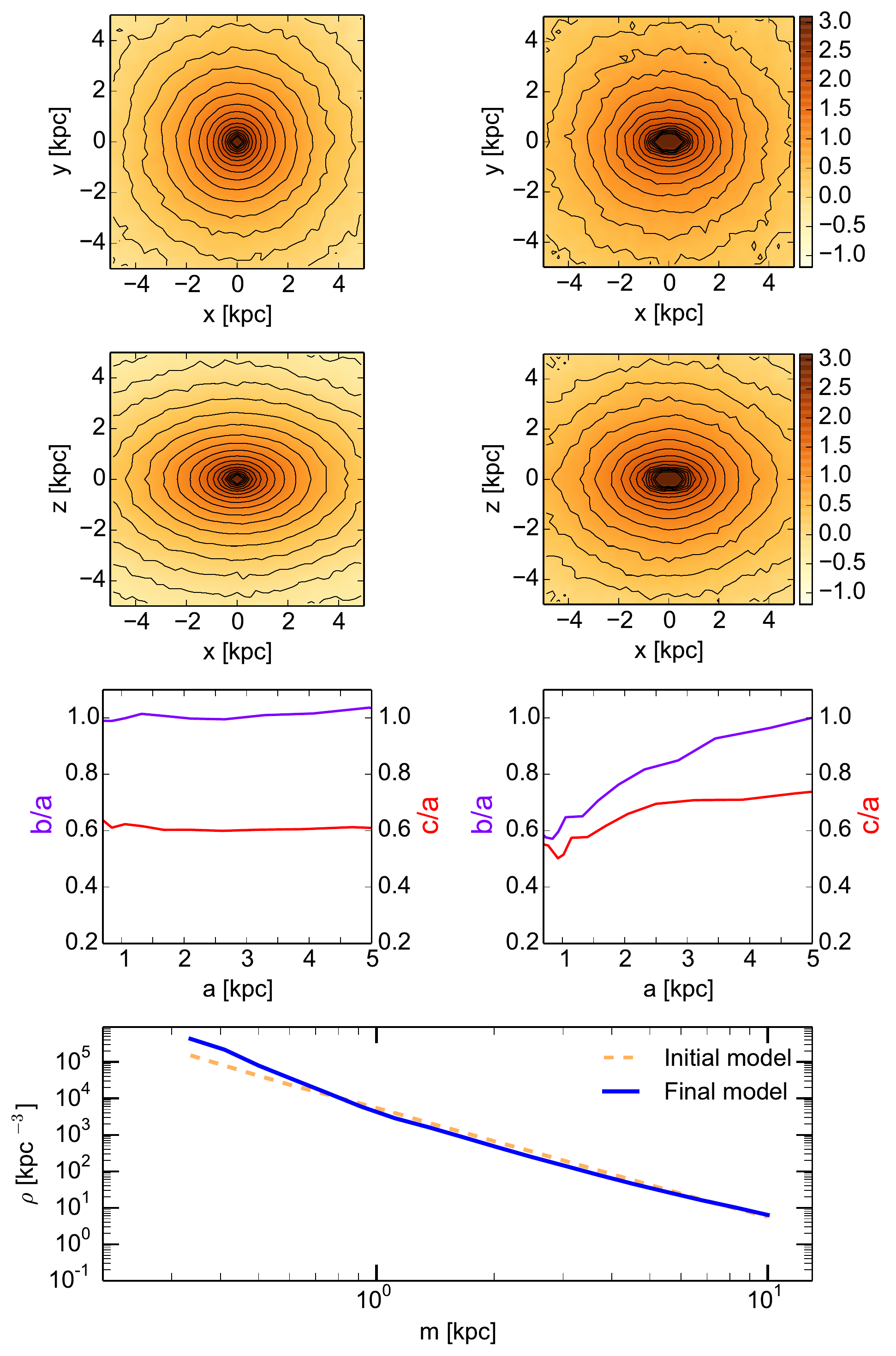}
\caption{Logarithmic surface density contours of the model stellar halo seen face-on (first row) and side-on (second
  row).  The third  row of panels shows the halo axis ratios b/a (purple line) and c/a (red line) as function of the major axis
  radius $a$.  The initial oblate distribution ($c/a=0.6$, left) has changed into a triaxial configuration with
  varying axis ratios at the final time (right). The bottom panel shows the density profile of the model at the initial and final times.}
\label{fig:shape}
\end{figure}

Figure \ref{fig:shape} shows that during the evolution, the shape of the stellar halo in the inner $5$ kpc of the Galaxy
has changed from the initial oblate distribution with constant axial ratio (left panels) to a triaxial distribution
(right panels). At the final time, the axial ratios of the triaxial stellar halo both increase with major axis radius,
from $\sim 0.6$ to $1.0$ in $b/a$, and from $\sim 0.5$ to $\sim0.7$ in $c/a$. The largest effects of the gravitational
influence of the Galactic bar and B/P bulge are seen in the central $3$ kpc. The final density profile, in oblate shells with flattening c/a=0.6, is similar to the initial stellar halo density as shown in Figure \ref{fig:shape}, although slightly steeper inside $\sim 0.6$ kpc.

\subsection {Kinematic Maps } \label {sec:kin_pred}

\begin{figure*}
\includegraphics[width=16cm]{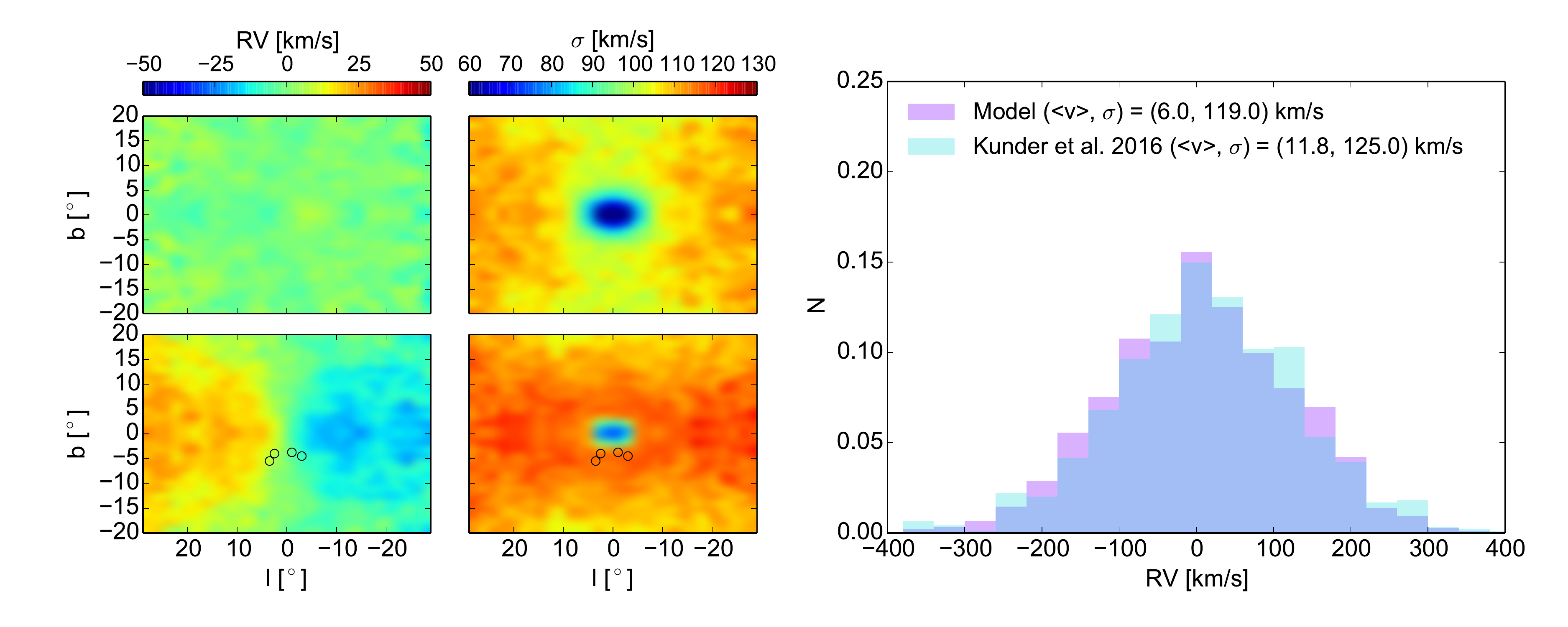}
\caption{ {\it Left:} Kinematic maps of the mean radial velocity (RV) and velocity dispersion at $t=0$ (top) and
  at the final time (bottom). {\it Right:} RV distribution in four bulge fields for RR Lyrae stars from
  \citet{Kunder2016} compared with our stellar halo model. The four fields are shown as circles in the lower kinematic
  maps on the left.}
\label{fig:kinematics}
\end{figure*}

We construct time-smoothed kinematic observables for the stellar halo model in the inner Galaxy \citep[see][]{Portail15}.  Figure
\ref{fig:kinematics} shows the predicted kinematics maps for the two snapshots $t=0$ (top panels) and at the final
time (bottom panels).  Before bar formation, the mean RV is approximately zero (the model does not initially rotate),
and the velocity dispersion is in the range $\sim70$ to $110$ km/s, depending on galactic longitude and latitude,
$(l,b)$. At the final time, the mean RV is $\sim15-25$ km/s at $\vert l \vert = 10\deg - 30\deg$ and $\vert b
\vert \leq 10\deg$, and the dispersion is $\sim120$ km/s outside the central few degrees.  The measured rotation in the
later stages of the N-body simulation is due to angular momentum transfer mostly during bar formation \citep{Saha12}.
Note that exact values are model-dependent, however, the main point is that the rotation is small compared to that of
the B/P bulge.

New kinematic data of the RRLs from \citet{Kunder2016} show that the RR Lyrae stars in the Galactic bulge at
$\vert l \vert =4\deg$, $b \sim -4\deg$ exhibit hot kinematics with almost negligible rotation. In the right panel of Figure
\ref{fig:kinematics}, we present the radial velocity distribution of our model versus their data, showing that although
the predicted rotation is $\sim 6$ km/s smaller, the RV histogram is in good agreement with the observations.

On the other hand, the metal-poor stars ($[\rm{Fe/H}]<-1.0$) in the ARGOS survey rotate faster \citep{Ness2013b} than
both our halo model and the bulge RRLs. This could be caused by a contamination of the ARGOS metal-poor component with
metal-poor thick disk stars, or by a larger initial rotation of the ARGOS metal-poor stars (as we have checked by
suitable models), but in any case the metal-poor halo component that we are discussing here must be different from the ARGOS
metal-poor stars.

\section {Orbital Analysis} \label{sec:orbits}

Our stellar halo model has a bar shape in the inner Galaxy. However, the rotation of this component is slow compared
with stars in the Galactic bar and B/P bulge. To understand the kinematic differences between both structures we analyze
the halo orbits.  To do so, we use frequency analysis. First, we compute the Fast Fourier Transform (FFT) for particle
tracks in major axis coordinate $x$ and cylindrical radius $R$, in order to identify the respective main
frequencies. Then we separate the stellar halo particles into bar-following particles, for which the ratio of
$f_R/f_x = 2\pm 0.1$, and not-bar-following particles, with $f_R/f_X\neq 2 \pm 0.1$. We are using the same criteria
to classify orbits as \citet{Portail2015b}.

\begin{figure*}
\includegraphics[width=15cm]{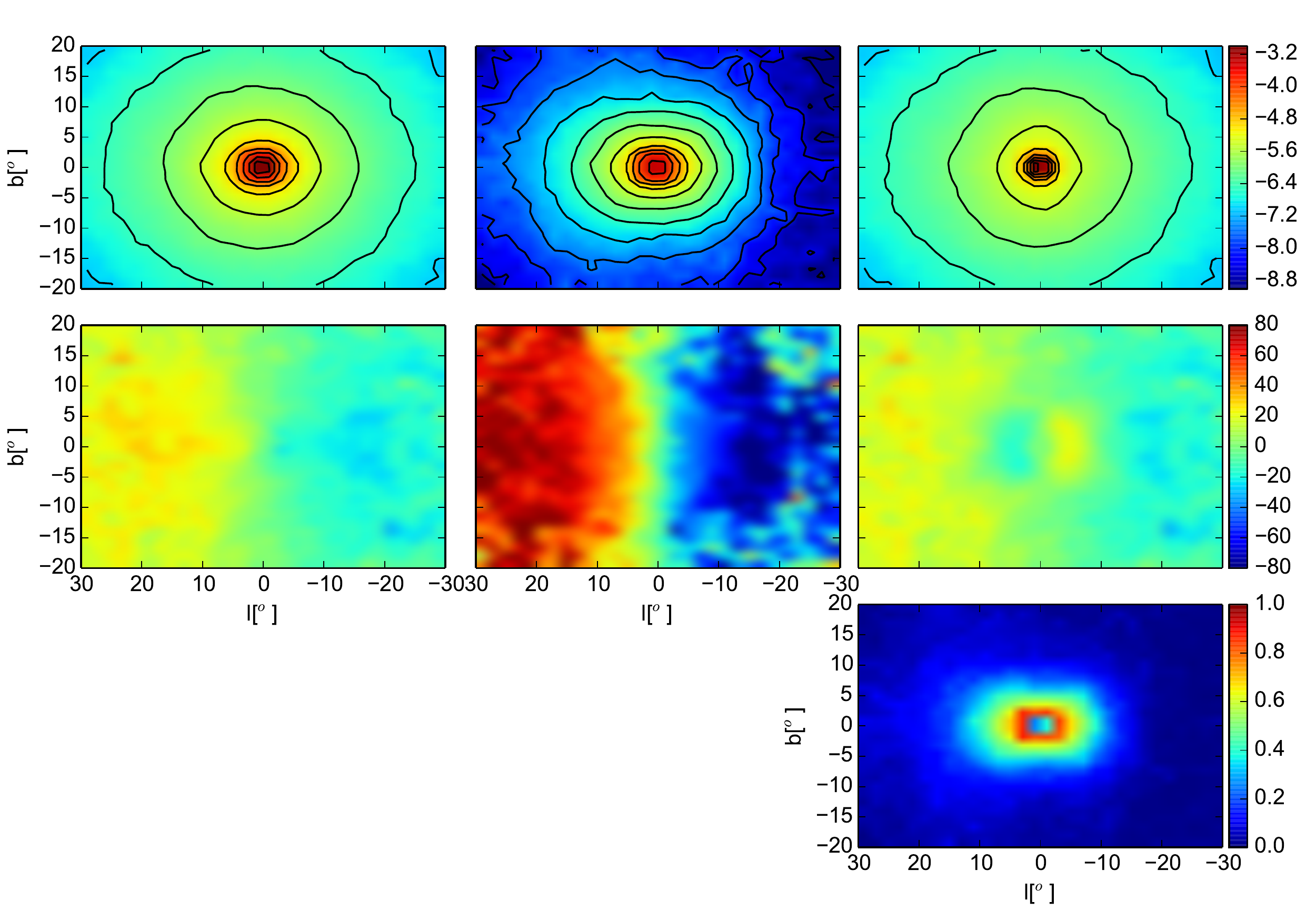}
\caption{{\it Top:} Surface density distribution for the entire sample of $30000$ stellar halo particles, for the
  bar-following halo particles, and the not-bar-following particles, from left to right (logarithmic scale). {\it
    Middle:} Radial velocity map for the entire sample, for the bar-following particles, and for the not bar-following
  particles, from left to right (in km/s).  {\it Bottom:} Relative contribution of the bar-following halo particles.}
\label{fig:orbits}
\end{figure*}

For a sample of 30000 stellar halo particles within $5$ kpc radius, we find that just $12\%$ of the halo particles
follow the bar and $88\%$ do not. In Figure \ref{fig:orbits}, we show the surface density and RV maps on the sky for the
entire sample, the bar-following particles, and the not-bar-following particles.  Notice that bar-following particles
have faster rotation ($\sim 80$ km/s) than the not-bar-following particles, but also that the relative contribution of
the bar-following particles is concentrated in the inner bulge region and is very small elsewhere.  We also see in
Figure \ref{fig:orbits} a counterrotating structure for the not-bar-following particles (third middle panel). A possible
reason for the presence of this structure is that the bar traps prograde orbits more easily than retrograde orbits.

\begin{figure}
\includegraphics[width=9cm]{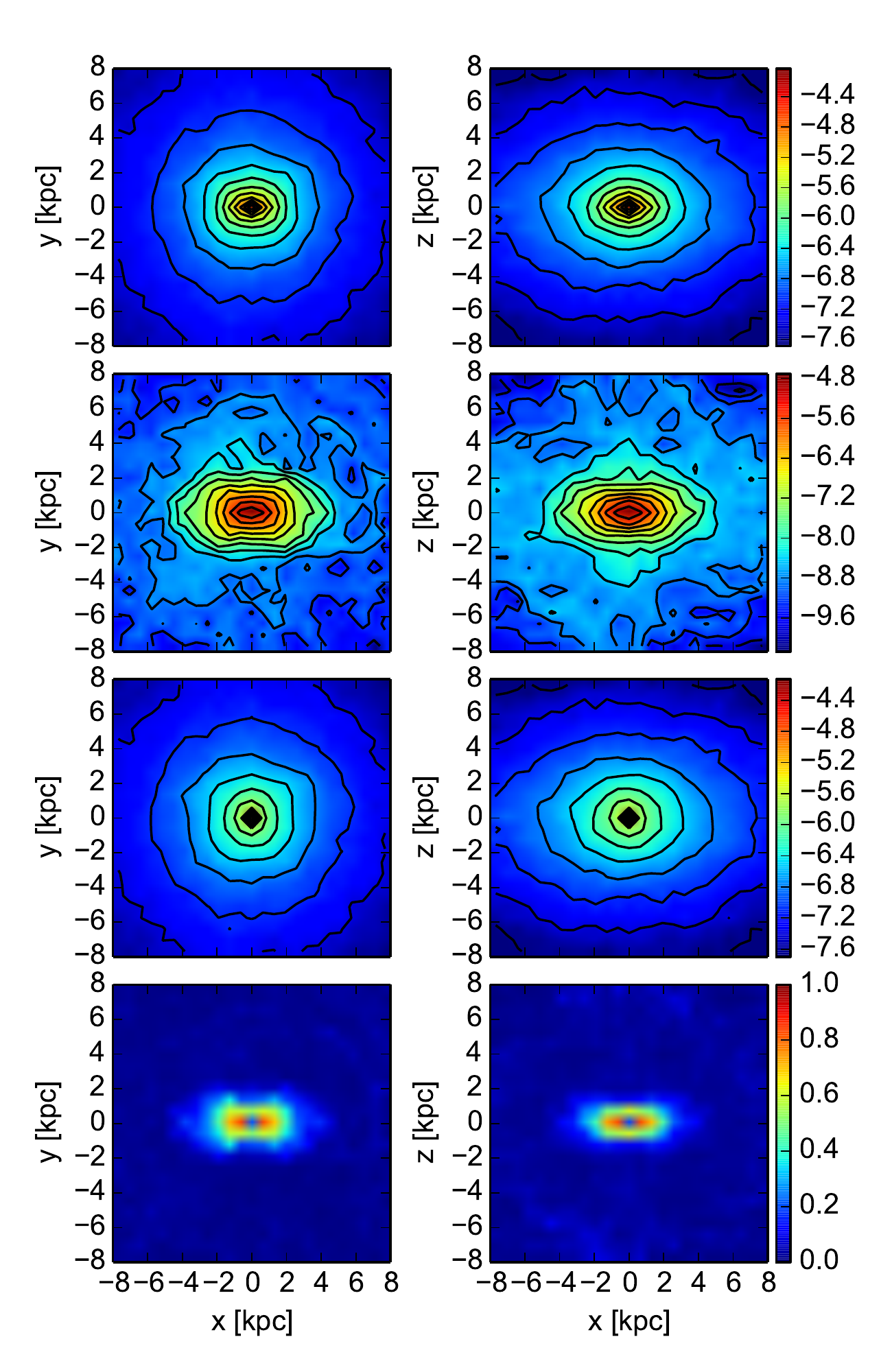}
\caption{Face-on (left panels) and side-on (right panels) surface density projections for the entire sample of $30000$
  (first row), the bar-following particles (second row), and the not-bar-following particles (third row; logarithmic
  scale). The bottom panels show the relative contributions of the bar-following halo particles in the face-on and
  side-on projections.}
\label{fig:orbits_xyz}
\end{figure}

Figure \ref{fig:orbits_xyz} is similar to Figure \ref{fig:orbits} but shows $(x,y)$ and $(x,z)$ projections rather than
$(l,b)$. With this Figure we point out that the bar-following particles do not show an X-shaped structure as is observed
for stars in the B/P bulge, and that the bar-following orbits dominate only in the inner few degrees where even their
rotation is small, and that their contribution elsewhere is also small.

\section{Conclusions}\label{sec:conclusions}

We constructed a model for the old stellar halo in the Milky Way (MW) as traced by RR Lyrae (RRL) stars in order to
investigate the origin of the triaxial shape and slow rotation of RRL in the bulge region.  Based on observational
constraints, we assumed an oblate shape with a single power-law density ($\rho\propto m^{-3}$) at early times. Using an
N-body simulation for the MW, we followed the evolution of this stellar halo component until several Gyr after formation
of the Galactic bar and B/P bulge. With this model, we addressed the effect of the bar and B/P bulge on the MW's inner
metal-poor halo, and made predictions for its density distribution and kinematics.  Our main conclusions can be
summarized as follows:
\begin{itemize}
\item [$\bullet$] The extrapolated density profile of RRLs in the bulge, with power-law index of $\simeq-3$,
  approximately agrees with the RRL density profiles in the stellar halo at $5-20$ kpc.
\item [$\bullet$] Through the gravitational influence of the Galactic bar and B/P bulge during their formation and
  evolution, the shape of the inner stellar halo evolves from oblate to triaxial.
\item [$\bullet$] Our model predicts a slow mean radial velocity in approximate agreement with recent measurements at
  $l\sim4\deg$, rising to $\sim15-25$ km/s at $l=10\deg-30\deg$, and a velocity dispersion of $\sim120$
  km/s. This rotation measured in the later stages of the N-body simulation is due to the angular momentum transfer during the bar evolution.
\item[$\bullet$] With frequency analysis we separated the stellar halo particles into bar-following and not-bar-following orbits, and found that bar-following orbits are a minority, $\sim12\%$ of all orbits within $5$ kpc. This is the reason for the slow rotation in our halo model.
  
\item [$\bullet$] The old component traced by RRLs in the bulge could be the inner extension of the Galactic stellar
  halo. It does not participate in the X-shape structure.

\end{itemize}

\section*{Acknowledgments}
We thank Christopher Wegg for helpful discussions.
APV acknowledges the support of a postdoctoral fellowship of CONACyT .

%%%%%%%%%%%%%%%%%%%%%%%%%%%%%%%%%%%%%%%%%%%%%%%%%%



%\bibliographystyle{mnras}


%%%%%%%%%%%%%%%%%%%%%%%%%%%%%%%%%%%%%%%%%%%%%%%%%%

%%%%%%%%%%%%%%%%% APPENDICES %%%%%%%%%%%%%%%%%%%%%

%\appendix



%%%%%%%%%%%%%%%%%%%%%%%%%%%%%%%%%%%%%%%%%%%%%%%%%%

% Don't change these lines
\bsp	% typesetting comment
\label{lastpage}
\end{document}